\documentclass[sigconf]{acmart}

\usepackage[english]{babel}
\usepackage[utf8x]{inputenc}
\usepackage[T1]{fontenc}

\setcopyright{none}
\acmDOI{}

\acmISBN{}

\acmConference[Accepted at ESEM'18]{[]Pre-print] Accepted at the International Symposium on Empirical Software Engineering and Measurement}{October 2018}{Oulu, Finland}
\acmYear{}
\acmBooktitle{Accepted at ESEM'18. This is a pre-print.}
\acmArticle{}
\acmPrice{}

\usepackage{amsmath}
\usepackage{graphicx}

\usepackage{booktabs} % For formal tables
\usepackage{todonotes}
\usepackage[caption=false,font=footnotesize]{subfig}
\usepackage{color}
\usepackage{multirow}
\usepackage{xspace}
\usepackage{natbib}
\usepackage{url}
\usepackage{paralist}

\newcommand{\SAP}{SAP\xspace}
\newcommand{\code}[1]{\textsf{\small#1\xspace}}

\long\def\longcaption#1#2{\centering\begin{minipage}{#1}\footnotesize\vspace{0.1\baselineskip}\noindent\emph{#2}\vspace{0.1\baselineskip}\end{minipage}}

\title{Vulnerable Open Source Dependencies: \\ \emph{Counting Those That Matter}}

\author{Ivan Pashchenko}
\affiliation{%
  \institution{University of Trento, IT}
}
\email{ivan.pashchenko@unitn.it}

\author{Henrik Plate}
\affiliation{%
  \institution{SAP Security Research, FR}
}
\email{henrik.plate@sap.com}

\author{Serena Elisa Ponta}
\affiliation{%
  \institution{SAP Security Research, FR}
}
\email{serena.ponta@sap.com}

\author{Antonino Sabetta}
\affiliation{%
  \institution{SAP Security Research, FR}
}
\email{antonino.sabetta@sap.com}

\author{Fabio Massacci}
\affiliation{%
  \institution{University of Trento, IT}
}
\email{fabio.massacci@unitn.it}

\renewcommand{\shortauthors}{Pashchenko et al.}

\begin{abstract}
\textbf{Background:}
Vulnerable dependencies are a known problem in today's open-source software ecosystems because OSS libraries are highly interconnected and developers do not always update their dependencies.
\textbf{Aims:}
In this paper we aim to present a precise methodology, that combines the code-based analysis of patches with information on build, test, update dates, and group extracted from the very code repository, and therefore, caters to the needs of industrial practice for correct allocation of development and audit resources.
\textbf{Method:}
To understand the industrial impact of the proposed methodology, we considered the 200 most popular OSS Java libraries used by SAP in its own software. Our analysis included 10905 distinct GAVs (group, artifact, version)
when considering all the library versions.
\textbf{Results:}
We found that about 20\% of the dependencies affected by a known vulnerability are not deployed, and therefore, they do not represent a danger to the analyzed library because they cannot be exploited in practice. Developers of the analyzed libraries are able to fix (and actually responsible for) 82\% of the deployed vulnerable dependencies.
The vast majority (81\%) of vulnerable dependencies may be fixed by simply updating to a new version, while 1\% of the vulnerable dependencies in our sample are halted, and therefore, potentially require a costly mitigation strategy.
\textbf{Conclusions:}
Our case study shows that the correct counting allows software development companies to receive actionable information about their library dependencies, and therefore, correctly allocate costly development and audit resources, which is spent inefficiently in case of distorted measurements. 
\end{abstract}
\keywords{Vulnerable Dependency, Open-Source Software, Mining Software Repositories}

\begin{document}
%%%%%%%%%%%%%%%%%%%%%%%%%%%%%%%%%%%%%%%%%%%%%%%%%%%%%%%%%%%%%%%%%%%%%%%%%%%%%
%%   UNCOMMENT THE FOLLOWING TO INCLUDE A FRONT-PAGE FOR ARXIV PRE-PRINT   %%
%%%%%%%%%%%%%%%%%%%%%%%%%%%%%%%%%%%%%%%%%%%%%%%%%%%%%%%%%%%%%%%%%%%%%%%%%%%%%
\begin{figure*}
\begin{minipage}{\textwidth}
\begin{center}
{\LARGE \bf Vulnerable Open Source Dependencies: Counting Those That Matter}\\[4mm]
{\Large [PRE-PRINT]} \\[8mm]
Ivan Pashchenko, Henrik Plate, Serena Elisa Ponta, Antonino Sabetta, Fabio Massacci\\
\end{center}

\vspace{10mm}
\noindent\textsc{Abstract}\\
\input{abstract}
\vspace{15mm}
\hrule
\vspace{10mm}
\begin{center}
{\Large Citing this paper}
\end{center}

This is a pre-print of the paper that appears in the proceedings of the \textbf{12th International Symposium on Empirical Software Engineering and Measurement}, 2018.

If you wish to cite this work, please refer to it as follows:\\[6mm]

\begin{verbatim}
@INPROCEEDINGS{pashchenko2018esem,
  author={Ivan Pashchenko and Henrik Plate and Serena Elisa Ponta and Antonino Sabetta and Fabio Massacci},
  booktitle={Proceedings of the 12th International Symposium on Empirical Software Engineering and Measurement (ESEM)},
  title={Vulnerable Open Source Dependencies: Counting Those That Matter},
  year={2018},
  month={Oct},
}
\end{verbatim}
\vspace{5mm}
\hrule

\end{minipage}
\end{figure*}
%%%%%%%%%%%%%%%%%%%%%%%%%%%%%%%%%%%%%%%%%%%%%%%%%%%%%%%%%%%%%%%%%%%%%%%%%%%%%
\maketitle

%====================================================================================================

\section{Introduction}
\label{sec:intro}
%===============================================================================
% use of OSS is widespread but there are known vulns
%===============================================================================
The inclusion of free open-source software (OSS) components in commercial
products is a consolidated practice in the software industry: as much as 80\% of the code of the
average commercial product comes from OSS~\citep{blackduck}.
SAP is an active user of and contributor to OSS\footnote{https://archive.sap.com/documents/docs/DOC-29056}. In this paper we report our hands-on experience on the industry relevant measurement of vulnerable dependencies in OSS.

Current dependency analysis methodologies are based on assumptions that are not valid in an industrial context. They may not distinguish dependency scopes~\cite{kula2017ese} (which may lead to reporting non-exploitable vulnerabilities), or consider only direct dependencies~\cite{cox2015icse} (although security issues may be introduced transitively~\cite{lauinger2017thou}). On the other hand, dependency analysis methodologies miss several important factors at all. For example, we could not find studies that distinguish dependencies, whose development had been suspended for unspecified time, although they may still introduce bugs and security vulnerabilities transitively. Additionally, current dependency management practices do not consider the fact that some dependencies are maintained and released simultaneously, and therefore, should be treated as a singular unit, while constructing dependency trees and reporting results of a dependency study.
%\as{processed: obscure to non-authors :-), be more explicit}
%together as a group.

%\as{this is about (2)}
%We could not find studies that distinguish dependencies, whose development had been suspended for unspecified time, although they may still introduce bugs and security vulnerabilities transitively. Additionally, current dependency management practices do not consider the fact, that some dependencies are maintained and released simultaneously, and therefore, should be processed
%%\as{processed: obscure to non-authors :-), be more explicit}
%together as a group.
% we do not find a study which approaches it in a holistic way. This motivated us to provide a systematized view on the aspects of the dependency management process that should be considered, while studying it: (i) results should not be reported for dependencies not included in a productive environment; (ii) libraries within same projects should not be considered as dependencies, since they are maintained and released simultaneously; (iii) dependencies, which development had been suspended for unspecified time, should be considered separately, since they may still introduce bugs and security vulnerabilities transitively; (iv) since developers typically treat the dependency update process as a low-priority task, the results should be free from falsely reported vulnerabilities that affect software dependencies.

These issues lead to 
% \as{revise}
% a misleading understanding of the situation regarding vulnerable dependencies and require software developers to spend additional effort to find and then fix vulnerable dependencies.
an inefficient allocation of costly development and audit resources due to the distorted measurements of vulnerable dependencies.

Hence, we make the following contributions:
\begin{compactitem}
\item A precise methodology, that caters to the needs of industrial practice, for reliable measurement of vulnerable dependencies in open-source software;
\item A tool to perform large-scale studies of (Maven-based) OSS libraries and to determine whether any of their dependencies are affected by known vulnerabilities
\item An empirical study of 10905 library instances of the 
200 Java Maven-based open-source libraries that are most frequently
used in SAP software.
\end{compactitem}

We found that as many as 20\% of the dependencies affected by a known vulnerability are \emph{not deployed}, and therefore, do not introduce vulnerabilities in the dependent library instances. Also, we found that the developers of the analyzed libraries could directly fix 82\% (45\% more comparing to a traditional approach) of their vulnerable dependencies. Our study indicates that, under a conservative model
to characterize halted dependencies, 14\% of the total number of dependencies are halted, and therefore, do not receive updates (including security fixes). Such dependencies should be used with caution, since mitigations of their vulnerabilities are costly.

\section{Terminology}
\label{sec:terminology}
In this paper we rely on the terminology established among practitioners and used in well-known dependency management tools such as Apache Ivy\footnote{\url{http://ant.apache.org/ivy/history/latest-milestone/ivyfile/dependency.html}} and Apache Maven\footnote{\url{https://maven.apache.org/pom.html\#Dependencies}}:
\begin{itemize}
\item A \textit{library} is a separately distributed software component, which typically consists of a 
logically grouped set of classes (objects) or methods (functions). To avoid any ambiguity, we refer
to a specific version of a library as a \textit{library instance}.
\item A \textit{dependency}\footnote{For the sake of consistency with the terminology used in Maven, we use the term `dependency' to denote a node (not an edge) of a dependency tree.} is a library instance, some functionality of which is used by another library instance (the \textit{dependent} library instance).
% \as{what does the following add?}
% \todo[inline]{IP: We define dependencies to be nodes in a dependency tree. Below I wanted to define what the edges are. Do you think, we can just skip the following?}\as{To me it just reads pedantic, with no substantial improvement of rigor or clarity. The only remark I would make is that while the term ``dependency'' would logically correspond to a relation, here, for the sake of consistency with the terminology used in Maven, that term is used to denote a node (not an edge) of the dependency tree.}
% The relation between a dependent library instance and its dependency may be described as follows: the dependent library instance \textit{depends on} its dependency, while dependency is \textit{used by} the dependent library instance.
\item A dependency is \textit{direct} if it is \textit{directly} invoked from the dependent library instance. 
% \todo{SP:I still do not understand the "dependent library" but i think it's too late :)}
\item A \textit{dependency tree}\footnote{We use the term \textit{dependency tree} rather than \textit{dependency graph} to be consistent with Maven, where the resolved graph of dependencies never contains cycles and each dependency appears once.} is a representation of a software library instance and its dependencies where 
each node is a library instance and edges connect dependent library instances to their direct dependencies.
\item A \emph{transitive dependency} is connected to the root library instance of a dependency tree through a path with more than one edge.
\item A \textit{project} is a set of libraries developed and/or maintained together by a group of developers. 
Dependencies belonging to the same project of the dependent library instance are \textit{own 
dependencies}, while library instances maintained by other projects are \textit{third-party 
dependencies}.
\item A \textit{deployed} dependency is actually delivered with the application or system that uses it,
while a \textit{non-deployed} dependency is only needed at the time of development (e.g., for testing)
but is not a part of the artifact that is eventually released and operated in a production environment.
%footnote{In this paper, we consider only the
%dependencies deployed by a software developer and not deployed from the platform itself.}
%\fm{This is either incomprehensible. It gives the idea that there is something we are hiding and we are %doing a very particular case. What is the platform? What are the libraries. Either we remove the note %altogther or we have to explain it fully. IF you are meaning the language libraries this is utterly %obvious}
\item A library instance is \textit{outdated} if there exists a more recent instance of this library at the time of analysis. A \textit{halted} library is such that the next estimated release time has been passed by far based on the interval of past releases (see \S\ref{sec:dead:alive} and \S\ref{sec:post-proc}).
\end{itemize}

\begin{figure}[t]
\vspace{-10pt}
\centering
\includegraphics[width=0.9\columnwidth, keepaspectratio]{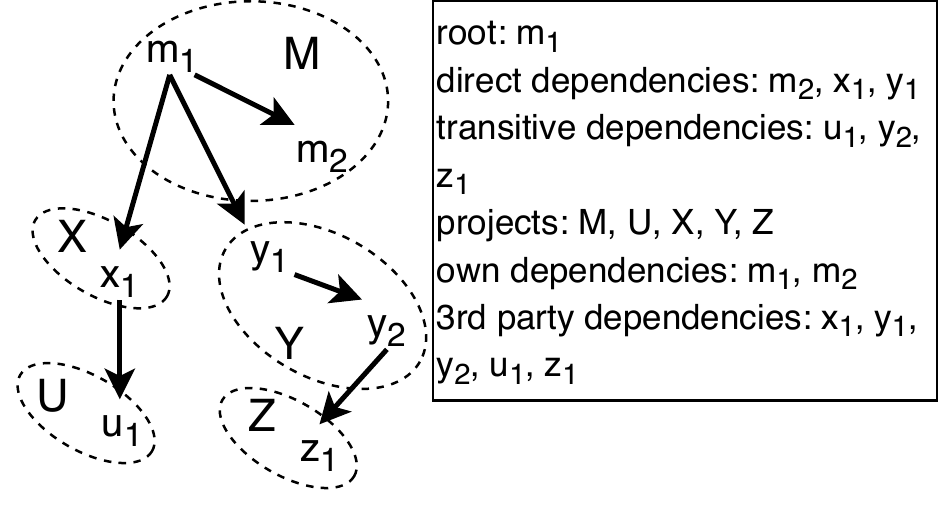}
\vspace{-10pt}
\caption{Dependency tree}
\label{fig:dep:grouping}
\vspace{-20pt}
\end{figure}

To illustrate how this terminology is used in practice, we refer to Figure~\ref{fig:dep:grouping}, which depicts the dependency tree for a library instance $m_1$. The library instance under analysis $m_1$ is the root, $m_2$, $x_1$, and $y_1$ are direct dependencies, while $u_1$, $y_2$, and $z_1$ are transitive dependencies. Library instances $m_1$, $m_2$ and $y_1$, $y_2$ are \emph{own dependencies} of projects $M$ and $Y$ respectively, while library instances $x_1$, $y_1$, $y_2$, $u_1$, and $z_1$ are \emph{third-party} dependencies of project M.
% for the root library instance $m_1$.

\section{Problem Statement}
\label{sec:problem}
The construction of the complete bill of materials (BoM) of a project is a necessary preliminary step to 
determining which dependencies of a project are vulnerable and assessing the risk they represent
and the effort needed to mitigate it.

Several approaches exist for analyzing software dependencies
(Table~\ref{table:relwork:rqs}). However, they do not provide a reliable measurement of the situation with software
dependencies, because they do not consider several key aspects:

\begin{compactitem}
\item a non-negligible number of dependencies that appear in the BoM
could not be possibly exploited because they are only used at development time
(e.g., for testing purposes) and are not delivered with the actual software system in operation;
\item libraries from the same project should be treated differently than third-party libraries
(the former should be maintained by the same team, which should fix them rather than wait
for another project team to release a new non-vulnerable version);
\item the mitigation strategy that should
be used to deal with each vulnerable library depends on the above two and on
the fact that the library might not be maintained any longer (\emph{halted}).
\end{compactitem}

\subsection*{RQ1: How many actually vulnerable dependencies does a library have?}
 \label{sec:sec:test:deploy}
A dependency tree for a library may include dependencies that are used only for testing or development purposes and are not deployed in the released version. Since they are not shipped with the product, they cannot possibly be exploited. Hence, allocating resources to fix or mitigate these vulnerabilities is pointless. This is well-known to software developers~\cite{kula2017ese}:
\begin{quote}
``\ldots In this case, it's a test dependency, so the vulnerability doesn't really apply \ldots''

``\ldots It's only a test scoped dependency which means that it's not a transitive dependency for users of XXX so there is no harm done \ldots''
\end{quote}

% Since \textit{non-deployed} dependencies are not shipped with the dependent library, any bugs and security vulnerabilities in those dependencies do not influence a project importing such a library.
% So the task of updating such dependencies has much lower priority comparing to the task of updating \textit{deployed} dependencies. The latter ones, in case of being vulnerable, may introduce bugs and software vulnerabilities into a dependent library.

Several recent works~\cite{kula2017ese,cox2015icse,cadariu2015tracking} do not mention explicitly that they consider only deployed dependencies (we discuss this further in Section~\ref{sec:relwork}). Indeed, the very quotes above in \cite{kula2017ese} show that the paper actually included such dependencies in its study. As a result vulnerable dependency count may become severely over-inflated.

%===========================================================================================
%===========================================================================================
\subsection*{RQ2: Who is responsible for vulnerable dependencies?}
\label{sec:sec:own:third}
A key question for the user of a vulnerable library is to attribute responsibility for fixing it (or avoiding projects with bad security discipline altogether). Developers of a software project are responsible for own code of their project and its direct dependencies (i.e., to keep them up-to-date). Although the concept seems intuitively simple, the following issues may occur:
\begin{compactitem}
\item \textit{Own vs third-party dependencies}: Failure to distinguish them may incorrectly present as an insecure ecosystem with several vulnerable dependencies (a ``dependency hell'' \cite{merkel2014docker}) what in reality is just a project that has broken its components into several libraries and did not fix its own vulnerable code.
\item \textit{Direct vs transitive dependencies}: A dependency tree may include several library instances that belong to the same project. Such dependencies
% are maintained and released simultaneously, they
should not be considered separately, since an update of one of those dependencies would automatically bring the new versions of all other dependencies from the same project. Hence, some transitive dependencies may actually be controlled directly from the project under analysis.
\end{compactitem}
%Indeed, dependency resolution tools (such as Maven, Gradle, etc.) rely on fine-grained dependency models that need to be correctly analyzed to draw meaningful conclusions about responsibility.
%Failure to group dependencies belonging to the same project, and

To illustrate these issues, we refer to the example of a dependency tree shown in Figure~\ref{fig:dep:grouping}. Both library instances $m_1$ and $m_2$ belong to the same project $M$. They are maintained and released simultaneously by the same team: if developers wanted to fix a bug in $m_2$, then they include the fix within the new release of the project $M$ and, at the same time, should update the versions of all their own libraries of project $M$ ($m_1$ and $m_2$). If they don't do so this might be a sign of a poor management within the project.

Suppose now that $m_2$, $y_2$, and $z_1$ are affected by known security vulnerabilities.
% The responsibility for fixing such vulnerabilities and the burden for doing\as{revise} can be misplaced if we do not distinguish between third-party and own dependencies.
\begin{compactitem}
\item Library instance $m_2$ is an \textit{own} dependency of $m_1$ because $m_1$ and $m_2$ belong to the same project $M$, and therefore, the source code of $m_2$ is under the control of developers of that project. Hence, the vulnerability should be fixed as part of the development of the project itself, i.e., by directly changing its source code. While from the perspective of the build system, $m_2$ is just a dependency, in practice it is a piece of vulnerable code developers are shipping as part of their project.
\item Dependency $y_2$ does not appear within configuration files of project $M$, but it comes together with $y_1$ (since both $y_1$ and $y_2$ belong to project $Y$), which, in turn, is a direct dependency of project $M$. Hence, developers of $M$ can control the version of $y_2$ by selecting a suitable $y_1$: if a newer version of $y_1$ is released, they should update project $M$ to use it.
\item Dependency $z_1$ appears within the dependencies of project $M$, since it is introduced transitively through project $Y$ (via $y_2$). Usage of dependency $z_1$ cannot be controlled without 
changing $M$ and transforming the (transitive) dependency $z_1$ into a direct dependency of the project. Since this would break the ``black-box'' dependency management principle, such a solution is not likely to be adopted. As a matter of fact, it is a responsibility of the developers of project $Y$ to keep the version of $z_1$ up-to-date.
\end{compactitem}

% Distinguishing these cases is very important for the correct allocation of development resources to fix possible issues that may be introduced into an industrial software library due to usage of OSS dependencies.
Proper distinction of these cases is very important for selection of an appropriate mitigation strategy and correct allocation of development resources for fixing security issues introduced by vulnerable dependencies.

%\as{\ldots?}

% As we mentioned above, the reported insecurity of the `dependency hell' (\cite{merkel2014docker}) might well and only be a presence of some insecure projects with many small fine-grained dependencies. 
%\todo[inline]{IP: Maybe we can find another term instead of "responsible for"}
%\textbf{RQ2}: \emph{Which part of vulnerable dependencies the developers of the analyzed libraries are actually responsible for?}

%================================================================================
%================================================================================
\subsection*{RQ3: How many direct dependencies can be actually fixed?}
\label{sec:dead:alive}

% Some dependency studies do not consider the origin of software dependencies and their relation to software projects, although this may present the actually manageable situation regarding vulnerable dependencies as an insecure `dependency hell' (\cite{merkel2014docker}). Instead, the library grouping by their belonging to software projects allows software developers to clearly understand which security vulnerabilities they may fix by directly updating their own library to use the fixed version of the dependency (the simplest mitigation strategy~\cite{reifer2003eight}), rather than considering an expensive mitigation for dealing with vulnerabilities in transitive dependencies.
If an outdated direct dependency is affected by a known vulnerability, the simplest solution to mitigate this vulnerability is to update the dependent library to use the fixed version of the dependency~\cite{reifer2003eight}. However, this becomes impossible, if an OSS library becomes inactive~\cite{kula2017ese}:

% Consider an example shown in Figure~\ref{fig:outdated:alive} which represent two possible temporal evolution for the dependency tree from Fig.~\ref{fig:dep:grouping}: at time $t_0$ the library instance $m_1$ depends on $y_2$, which, in turn, depends on $z_1$. Then a security vulnerability is discovered in version $v_1$ and fixed in version $v_2$ of dependency $z_1$. The developers of version $v_2$ of $y_2$ can adopt the safe version of $z_1$, and be free from vulnerable dependencies at the moment $t_1$. The developers of the root library $m_1$ could also update the latest available version of direct dependency $y_2$ to avoid the old dependency. They might not do it for functional reasons (See e.g.\ Figure~10 in \cite{dashevskyi2018screening} for the sheer impact of an update of Apache CXF) and might end up with an outdated dependencies as in Figure~\ref{fig:outdated:alive}. Still, they had a choice.

%---------------------------------------------------------------------------------------
% \begin{figure*}[!ht]
% \subfloat[All dependencies are ``alive'']{
% \includegraphics[width=\columnwidth, keepaspectratio]{graph/outdated_alive}
% \label{fig:outdated:alive}
% }
% %-------
% \subfloat[Depending on an halted dependency]{
% \includegraphics[width=\columnwidth, keepaspectratio]{graph/outdated_dead}
% \label{fig:outdated:dead}
% }
% \caption{``Alive'' and halted dependencies}
% \label{fig:outdated:alive:dead}
% \end{figure*}
%---------------------------------------------------------------------------------------
\begin{figure}
    \centering
    \includegraphics[width=0.8\columnwidth, keepaspectratio]{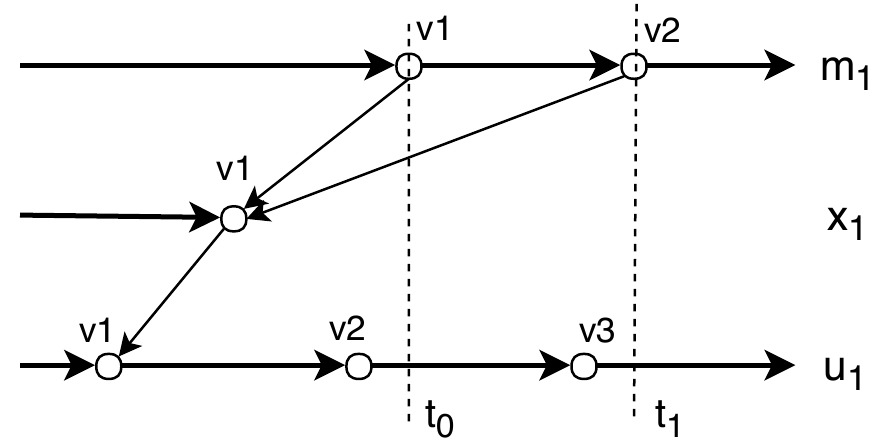}
    \longcaption{\columnwidth}{Library $m_1$ has a halted dependency $x_1$. In case a vulnerability is discovered in $x_1$ or its dependency $u_1$, there would be no version of $x_1$ that fixes such a vulnerability or adopts a fixed version of $u_1$.}
    \vspace{-10pt}
    \caption{Halted dependency}
    \label{fig:halted:dependency}
\vspace{-10pt}
\end{figure}
%---------------------------------------------------------------------------------------

% \cite{cox2015icse} recommends to keep direct dependencies up-to-date to decrease the probability of a project to be affected by bugs and security vulnerabilities. However, in case a certain project depends on a library, which has reached an end-of-maintenance point, keeping this library ``up-to-date'' may not save a project from having outdated transitive dependencies (with their bugs and security vulnerabilities).

% It may also happen that a OSS library becomes inactive~\cite{kula2017ese}:
\begin{quote}
``\ldots our project has been inactive and production has been halted for indefinite time''
\end{quote}

If a security vulnerability is discovered in a no longer actively developed library, there would be no version of this library that fixes the vulnerability. Hence, being a dependency, this library will introduce the vulnerability to all its dependents.
Additionally, a halted dependency may transitively introduce outdated dependencies and expose the root library instance to bugs and security vulnerabilities (Figure~\ref{fig:halted:dependency}): the root library instance $m_1$ depends on the last version of halted dependency $x_1$, which, in turn, uses an ``alive'' dependency $u_1$. Although both versions $v1$ and $v2$ of library $m_1$ technically use the latest available version of direct dependency $x_1$, there would always present outdated transitive dependency $u_1$. Hence, the halted dependencies should be considered separately.

Clearly, the presence of halted dependencies has a major impact on a company maintenance strategy. Indeed, any user of library $x_1$ would not obtain any benefit from switching to its latest version. The vulnerable version of $u_1$ would always be present. A different mitigation strategy might be needed: (i) contribute to the halted library, i.e., to develop its new release; or (ii) fork the halted library and continue its maintenance as part of the dependent library.

%================================================================================
%================================================================================
\subsection*{Observation: Name-Based vs Code-Based vulnerability matching}
\label{sec:vuln:matching}

The main source of vulnerabilities in software components is the 
National Vulnerability Database 
(NVD\footnote{\url{https://nvd.nist.gov/}}) that uses
the Common Platform Enumeration (CPE) standard for enumerating the 
affected components. The NVD represents the most 
complete, public source of vulnerabilities\footnote{Other sources of vulnerabilities 
are software-specific advisories and bug tracking systems which are 
used to report and solve security issues. Some of them might be product or vendor specific, e.g. MSFA for Mozilla's Firefox browser.} albeit it 
does not cover all OSS projects with the same 
accuracy. Moreover, CPE names, used to denote the affected 
software, use a different granularity and convention than software 
package repository coordinates. 

Most approaches for the identification of vulnerable dependencies use the NVD and try to map CPE names to the language-specific naming, e.g., Maven coordinates. This is, for example, the case for OWASP Dependency Check\footnote{\url{https://www.owasp.org/index.php/OWASP_Dependency_Check}}. Such approaches suffer from both false positives and false negatives. In particular, many false positives come out of the fact that CPE names are more coarse grained than Maven coordinates: a vulnerability only affecting the \code{poi-ooxml} artifact within the Apache Poi project, would be assigned to the entire project in the NVD, thereby resulting in false positives whenever an application only uses `Poi' artifacts other than \code{poi-ooxml}. This might be further exacerbated since the NVD might use an over-approximation rule 'X and all previous versions' for marking vulnerable versions (See, for example, \cite{nguyen2013reliability,nguyen2016automatic} for the study of browser vulnerabilities and the large presence of false positives).

% False negatives easily result from the fact that the NVD is not complete and whenever the assigned CPEs are not listing all required softwares (e.g., in some cases the NVD assigns vulnerabilities to products rather than the responsible libraries).

% Backward patch analysis \cite{plate2015impact,nguyen2016automatic,dashevskyi2018screening} has been proved to be particularly effective in tracking the vulnerable versions of past releases in OSS and we plan to rely on it to obtain the same result for a more precise report on vulnerable dependencies.

% \textbf{RQ4}: \emph{Can we design a system for the collection of vulnerable dependencies by patch analysis of the libraries of interest with minimal reliance on vulnerability databases?}

% To address RQ4 we will base our entire data collection on code analysis (See Section \ref{sec:vuln:identification}).

\section{Methodology}
\label{sec:methodology}
The methodology that we present in this section is language independent and it
only relies on the availability of a dependency management mechanism, such as those provided
for Java (Maven, Gradle), Javascript (npm), Python (pip), PHP (pear), and so forth.

Considering the popularity and industrial relevance of Java\footnote{Java is estimated to
be the most popular programming language since 2004, according to the two indexes used by IEEE Spectrum (\url{http://spectrum.ieee.org/}) to assess popularity of a programming language: (i) Tiobe index (\url{http://www.tiobe.com/tiobe-index/}), which combines data about search queries from 25 most popular websites of Alexa; and (ii) PYPL index (\url{http://pypl.github.io/PYPL.html}), which uses Google search queries.}, in the following we demonstrate our methodology on Java projects.

Over the past decade,
Apache Maven established itself as a standard solution in the Java ecosystem for dependency
management and other tasks related to build processes.
Other solutions exist, such as Apache Ivy~\footnote{\url{http://ant.apache.org/ivy/}}
and Gradle (which is gaining popularity)\footnote{\url{https://gradle.org/}},
however Maven\footnote{\url{https://maven.apache.org/}} still has the largest
share of users\footnote{\url{https://zeroturnaround.com/rebellabs/java-tools-and-technologies-landscape-2016/}}. Hence, we use it to demonstrate the proposed mitigations for each problem described in Section~\ref{sec:problem}, although the concepts presented below can be easily extended to other dependency management systems.
%\todo[inline]{IP: Can you, please, check if the text above fixes the problem of the study scope?}

In Maven the name of a component is standardized\footnote{\url{https://maven.apache.org/guides/mini/guide-naming-conventions.html}} and represented as \textit{groupId}:\textit{artifactId}:\textit{version}. Hence:
\begin{compactitem}
\item a ``project'' may be referenced as Maven \textit{groupId}
\item a ``library'' corresponds to \textit{groupId}:\textit{artifactId} (GA)
\item a ``library instance'' corresponds to the name of Maven component \textit{groupId}:\textit{artifactId}:\textit{version} (GAV)
\end{compactitem}

% In Sections \ref{sec:dep:resolution} - \ref{sec:vuln:identification} we describe the steps of the methodology that we propose to follow while analyzing vulnerable dependencies in OSS. In Section~\ref{sec:tool} we show how the proposed methodology could be wrapped up with a tool.

%========================================================================================
\subsection{Dependency resolution}
\label{sec:dep:resolution}

For each of the library instances in our sample, we use Maven to determine the complete set
of dependencies. Before doing so, Maven requires that the Project Object Model (POM) files be installed in the
local repository.
Once a POM is installed locally, we use the standard Maven goals\footnote{In Maven terminology, \emph{goal}
can be thought of as a synonym of a \emph{command}.} \textit{dependency:tree} and
\textit{dependency:resolve} to construct the dependency tree of each POM and to resolve conflicts
and duplications.

%========================================================================================
\subsection{Post-processing of the results}
\label{sec:post-proc}

The next step of our data collection process is a post-processing of the data obtained after the dependency resolution step to address the problems discussed in Section \ref{sec:problem}.

\noindent\textbf{Filter non-deployed dependencies.}
To control whether a dependency is deployed with an artifact, Maven provides a possibility for a software developer to specify the \textit{scope}. The dependencies with scopes \textit{provided} and \textit{test} are used only for development purposes and are not shipped with a released artifact, hence, we do not consider them for the further analysis as non-deployed dependencies.

\noindent\textbf{Dependency grouping.}
The Maven dependency resolution process
starts from the \textit{POM}s under analysis as the major source of the necessary information to build the dependency trees. However, at the final step of our analysis, the vulnerable dependency represents the most valuable asset. Hence, we perform the final aggregation of the results in the opposite direction, i.e., considering the paths from vulnerable dependencies to libraries under analysis:
\begin{compactitem}
% \item we identify the first GAV in the path with the \textit{groupId} of an analyzed GAV and ``cut'' the path after this GAV, i.e., we do not analyze software dependencies within the same project
\item we group all GAVs with the same groupId within one path and substitute them in the path with the GAV, closest to the vulnerable GAV
\end{compactitem}

Consider the example of a dependency tree from Figure \ref{fig:dep:grouping}: let dependencies $x_1$ and $z_1$ be affected by known security vulnerabilities. Initially there are two paths from vulnerable dependencies to the analyzed root library: ($x_1$, $m_1$) and ($z_1$, $y_2$, $y_1$, $m_1$). In the second path library instances $y_1$ and $y_2$ belong to the same project $Y$, hence, they are grouped. So, the analysis results in two vulnerable paths: ($x_1$, $m_1$) and ($z_1$, $y_2$, $m_1$).

\noindent\textbf{Identification of halted dependencies.}
Public software package repositories keep all published library instances, since there is a possibility to break a build of a library in case of a library dependency is removed~\cite{kikas2017structure}. So, even a certain library is no longer maintained, it would still be available from a software package repository.

At the same time, when selecting a mitigation strategy, software developers need to know that a fixed version of a vulnerable dependency is going to appear (otherwise, a costly mitigation strategy is required).
Some projects publish information about their decision to stop maintenance of certain libraries. Monitoring these sources of information requires tracking notifications for every dependency separately; some projects do not publish such data frequently, or stop publishing it at all. At this time, there is no systematic and scalable approach to determine if an OSS component has reached the end of its lifetime.

We propose to consider the amount of time library developers require to release a new version for determining whether a library development becomes halted.

Some libraries may have varying time intervals between releases due to different release strategies adopted within development teams, as well as the maturity of a certain library:
% Library maturity changes over time:
at earlier stages of development it needs to have more updates than an established library with a long development history.
An example of a mature library is the Apache \code{commons-logging} package. Released on 2007-11-26 version 1.1.1 was the latest available version for more than 5 years till the release of version 1.1.2 on 2013-03-16.
% Therefore, for each individual library we compute the ``\textit{Last release interval}'', that library developers need to release a new version.

% A new library may be released every couple of weeks, while on later stages the same library may be released once in a month or even years.
Since the time difference between recent releases should have bigger impact on the \textit{Last release interval} comparing to the time difference between older releases, the typical statistical model that describes such a process is a simple Exponential Smoothing model~\cite{brown1959statistical}:
\begin{equation*}
  \begin{aligned}
  \text{Last release interval} & = & \alpha \sum^{n}_{i=0} \left \{(1 - \alpha)^i * \text{Release time}_{n - i} \right \} \\
  \text{Expected release date} & = & \text{Last release} + \text{Last release interval}
  \end{aligned}
\end{equation*}

\noindent where $\text{Release time}_{i}$ is the time needed to release the $i$-th version of a library, $0<\alpha<1$ is the smoothing parameter that shows how fast the influence of previous time intervals decreases\footnote{Our hands-on experience (which is also supported by the observation of released dates for the analyzed libraries) suggest, that the last three releases have the major impact on the \textit{Expected release date} of a library, and therefore, in this paper we count $\alpha = 0.6$. For libraries with less than 3 releases, we take the \textit{Last release interval} equal 3 months.}.
We estimate the \textit{Expected release date} for a library by adding the \textit{Last release interval} to the release date of the latest available version of the library. Then we determine the status of the library as follows:
\begin{compactitem}
\item \textit{next release date} $<$ \textit{TIME}: the library is \textit{halted}
\item \textit{next release date} $\geq$ \textit{TIME}: the library is \textit{alive}
\end{compactitem}

\textit{TIME} represents the date, for which the library status is calculated. In our study, for each analyzed library instance we will identify its release date and use it to calculate whether any of the dependencies were halted. To know the current status, \textit{TIME} should be equal to the current date. 

% \as{If I were a reviewer I would object that the halting model used here is not validated and it is not clear if it actually fits real projects; can we say that the value of $\alpha$ we chose is conservative (is it?) and therefore it is obvious that, despite we cannot precisely quantify how bad the problem is, there is obviously a problem when halted dependencies are not distinguished as such? And that the problem is clearly there regardless of what model one uses to characterize halted dependencies? At the very least, can we say that we did check, for each dependency that our model determined to be halted, that it was (or appeared to be) indeed halted by observing its repository, the project web page, or other resources on the web?}

The proposed model based on release dates is conservative, since it provides the lower bound for the estimation of the \textit{Expected release date} for a library. Hence, it is more likely to be affected by false positives, i.e., to classify a library as halted, when it is still under development. However, such finding would mean, that a library does not receive a fix for a long period of time, which increases chances of a zero day vulnerability to be exploited. Hence, even in case of ``false positives'', our model provides a valuable information for a software developer.

To examine the reliability of the proposed model, we randomly selected 100 distinct library instances identified to be halted. Then we manually looked for any available information of whether a new version of a halted library is planned to be released. For this purpose, for every halted library we checked (when possible) (i) their software repositories, (ii) release pages, or (iii) other available resources returned by Google searches. The manual analysis did not reveal any libraries falsely reported to be halted.

% The model we propose for identification of halted dependencies may sometimes introduce misclassifications for libraries that become inactive for unexpectedly long periods of time. However, in this study we do not aim to create a precise model for determining halted libraries, but rather to raise awareness about threats of depending on them. Moreover, 
% our methodology can be easily adjusted to work with a more precise model for determining a status of a library.

%========================================================================================
\subsection{Identification of vulnerabilities}
\label{sec:vuln:identification}

Once all the dependencies of each subject project are determined,
we lookup any known vulnerabilities associated with them.
As we mentioned, an obvious implementation relies on a database that lists the known security vulnerabilities and the exact library instances that are affected.
Unfortunately, this information is not readily available from standard sources, such as the NVD, where vulnerabilities are assigned
to components that are designated at a much coarser level of granularity than we need.

To improve our precision we
leverage on code based approaches to vulnerability detection such as Ponta et al.~\cite{ponta2018beyond} and Dashevskyi et al.~\cite{dashevskyi2018screening}. 
Starting from known vulnerabilities from the NVD, advisories, bug tracking systems, etc., the commit fixing the vulnerability is identified manually and analyzed resulting in a list of code changes. All software constructs (e.g., constructors, methods) included in such list are the so-called \emph{vulnerable code}. The creation of such knowledge is a one-time effort for each vulnerability. 
Then, for every analyzed project, the list of all own libraries of the project and all its dependencies is collected by performing a code-level matching of the vulnerable fragment following the approach of~\cite{ponta2018beyond}.
% \as{Is this really what was done? Unless I'm missing something, we used Vulas to check each one of the library instances...}
% \ip{I have applied the fix to this and the following comment, as suggested by Fabio}
Whenever the vulnerable code fragment is contained within a dependency, the corresponding vulnerability is automatically reported for our analysis. Additional
details on the procedure are reported in~\cite{ponta2018beyond}.
% \as{here too: when Vulas reported that the dependency had issues, the issues were reported...; anyway, \cite{ponta2018beyond} or~\cite{dashevskyi2018screening}, this part would read like a lot of hand-waving to anyone not familiar with those works, maybe a couple of paragraphs should be devoted to a summary?}

% Since Dashevskyi et al.~\cite{dashevskyi2018screening} use an approximate slicing method, there might still be problem of false positives (as well as false negatives) for which the paper provides also some statistical confidence interval by using Agresti-Coull methods. To remove part of the uncertainty, we integrate such information with static and dynamic analysis techniques as advocated by Plate et al.~\cite{plate2015impact}.
% %%%%%%%%%%%%% TODO %%%%%%%%%%%%%%%%%%%%%%%%%%%
% \as{I think this reference to plate2015impact is not correct: static/dynamic analysis was
% not used in our experiment! what if we stop at ``\ldots Agresti-Coull methods''? }
%%%%%%%%%%%%% TODO %%%%%%%%%%%%%%%%%%%%%%%%%%%
% \todo{SP: Unfinished sentence, delete?}
%========================================================================================
\subsection{Dependency resolution tool}
\label{sec:tool}

To automate our dependency study we implemented a tool that:
\begin{compactitem}
\item wraps \textit{dependency:tree} and \textit{dependency:resolve} Maven commands, which helps us get
a more manageable (and a machine-readable) representation of the results of the resolution mechanism. This allows us to construct the resolved dependency tree for each analyzed library instance.
\item uses the code-based approach of~\cite{ponta2018beyond} to annotate dependency trees with the vulnerability data at our disposal. In particular, when a vulnerable library instance is found among the dependencies of one of the analyzed root libraries,
our tool produces in the output (i) the identifier of the vulnerability, (ii) the library instance importing it,
and (iii) the complete dependency path leading from the root library to the vulnerable dependency.
\item applies path simplifications and produces the results in the form of a human-readable report.
\end{compactitem}

\section{Data collection}
\label{sec:data}
\begin{table}[t]
\centering
\caption{Descriptive statistics of the library sample}
\label{table:stats}
\longcaption{\columnwidth}{We considered the 200 most popular OSS Java libraries used by SAP in its own software, which resulted in 10905 distinct GAVs when considering all library versions.}
\footnotesize
% \begin{tabular}{|c|c|c|c|c|c|c|}
% \hline
%       & \#libs & \#GAVs & \#deps                         & \begin{tabular}[c]{@{}c@{}}\#direct\\ deps\end{tabular} & \begin{tabular}[c]{@{}c@{}}\#transitive\\ deps\end{tabular} & \#usages                        \\ \hline
% Total & 200         & 10 905 & 129 655                        & 46 458                                                  & 83 197                                                      & 610 229                         \\ \hline
% Avg.  & 1           & 54,53  & 11,89 $\pm$ 18.54 & 4,26 $\pm$ 6,80                            & 7,63 $\pm$ 13,56                               & 55,96 $\pm$ 508,41 \\ \hline
% \end{tabular}
\begin{tabular}{|c|c|c|c|c|c|c|c|}
\hline
                  & $mu$  & median & $sigma$ & min & max    & Q1   & Q3   \\ \hline
\#GAVs            & 54.52 & 35.0   & 49.24   & 1.0 & 248    & 15.0 & 87.0 \\ \hline
\#dependencies    & 11.89 & 3.0    & 18.54   & 0.0 & 131    & 0.0  & 16.0 \\ \hline
\#direct deps     & 4.26  & 2.0    & 6.80    & 0.0 & 51     & 0.0  & 6.0  \\ \hline
\#transitive deps & 7.63  & 1.0    & 13.56   & 0.0 & 92     & 0.0  & 11.0 \\ \hline
\#usages          & 55.96 & 5.0    & 508.41  & 1.0 & 29 472 & 1.0  & 23.0 \\ \hline
\end{tabular}
\vspace{-10pt}
\end{table}

Processing of a full Maven Central repository with almost 2,7 million GAVs would be impractical
and especially would include artifacts of no relevance in industrial practice.
Hence, for this paper we take a sample from Maven Central, as explained below.

\noindent\textbf{Library selection - incorrect way}. Initially,
% to determine the popularity of a certain library instance,
we followed the approach of~\cite{sajnani2014popularity} and selected the number of usages of a library instance as a proxy for its popularity. By usage we understood the number of direct dependent library instances of a library instance of interest\footnote{We used the data from MVNrepository (\url{https://mvnrepository.com/}).}. 

However, when we extracted the list of top 100 most used libraries, the resulted list had an unbalanced usage distribution: \code{scala} and \code{spring-framework} projects were over-represented, while some well-known projects, like Apache Tomcat, were not present in the list. A possible reason may be in the large difference in numbers of own libraries in different projects: if a project has 100 own libraries and they directly depend on a certain library instance, then this library instance would be ``used'' 100 times, while in reality there is only one usage.

This approach may have potentially allowed us to receive a ``good'' list of libraries, if as a proxy for popularity we used the number of dependent projects. However, such information is not easily available (to obtain it, we would have to build dependency trees for all GAVs in Maven Central), so we had to find another way to construct the list of libraries for our study.

\noindent\textbf{Library selection - the way we followed}.
To ensure industrial relevance of our study, we selected the top 200 OSS libraries used by a set of more than 500 Java projects developed at SAP; these include actual SAP products and software developed by the company for internal use. Those libraries comprise, for instance, \code{org.slf4j:slf4j-api} and \code{org.apache.httpcomponents:httpclient}, and correspond to 10905 library instances when considering all versions (see Table \ref{table:stats} for
% \as{here and elsewhere: description statistics $\rightarrow$ {\bf descriptive} statistics}
descriptive statistics of the selected sample).

\section{Results and Discussion}
\label{sec:results}
% \textcolor{blue}{
% org.apache.logging.log4j:log4j-core:jar:2.9.1 depends on org.codehaus.plexus:plexus-utils:3.1.0:test (Vulnerability CVE-2017-1000487 - Shell Command Injection, CVSS v.3.0 Severity 9.8 Critical)
% }

In this Section we answer the research questions and present how each step in our methodology influences the results of a dependency study for the complete sample of selected libraries. Then we show the impact of the proposed methodology on the results of a dependency analysis for an average industrial software library. Below we present the results from the perspective of developers of the analyzed libraries.

% \subsection*{Effect of the proposed methodology on the sample of analyzed libraries}

\paragraph{RQ1: How many actually vulnerable dependencies does a library have?}

%---------------------------------------------------------------------------------------
\begin{table*}[t!]
\centering
\caption{The effects of considering only deployed dependencies (RQ1) and grouping dependencies by software projects (RQ2)}
\label{table:cure:effect}
\longcaption{\textwidth}{The left part of the table shows the number of vulnerabilities within all and deployed dependencies. Non-deployed dependencies represent significant part within both vulnerable (45\% of direct and 34\% of transitive) and non-vulnerable (20\% direct and 22\% transitive) dependencies. The right part of the table shows the effect of the grouping step, which allowed us to reveal that developers of analyzed libraries could directly fix 82\% of deployed vulnerable dependencies (direct vulnerable dependencies). Moreover, dependency trees of 148 analyzed library instances included own dependencies. Although they are not affected from known vulnerabilities, they may still introduce some noise, while planning allocation of development resources.}
% \footnotesize
\begin{tabular}{c|c|c|c|c|cc|c|c|c|c|c|c|c|c|}
\cline{2-5} \cline{8-15}
\multirow{2}{*}{}              & \multicolumn{2}{c|}{Not vuln}                                                                                 & \multicolumn{2}{c|}{Vuln}                                                                                     &                       & \multirow{2}{*}{} & \multicolumn{2}{c|}{Not vuln}                                                                                 & \multicolumn{2}{c|}{Vuln}                                                                                     & \multicolumn{2}{c|}{Not vuln}                                                                                & \multicolumn{2}{c|}{Vuln}                                                                                    \\ \cline{2-5} \cline{8-15} 
                               & \begin{tabular}[c]{@{}c@{}}Direct\\ dep.\end{tabular} & \begin{tabular}[c]{@{}c@{}}Trans.\\ dep.\end{tabular} & \begin{tabular}[c]{@{}c@{}}Direct\\ dep.\end{tabular} & \begin{tabular}[c]{@{}c@{}}Trans.\\ dep.\end{tabular} &                       &                   & \begin{tabular}[c]{@{}c@{}}Direct\\ dep.\end{tabular} & \begin{tabular}[c]{@{}c@{}}Trans.\\ dep.\end{tabular} & \begin{tabular}[c]{@{}c@{}}Direct\\ dep.\end{tabular} & \begin{tabular}[c]{@{}c@{}}Trans.\\ dep.\end{tabular} & \begin{tabular}[c]{@{}c@{}}Own\\ lib.\end{tabular} & \begin{tabular}[c]{@{}c@{}}3rdParty\\ dep.\end{tabular} & \begin{tabular}[c]{@{}c@{}}Own\\ lib.\end{tabular} & \begin{tabular}[c]{@{}c@{}}3rdParty\\ dep.\end{tabular} \\ \cline{1-5} \cline{7-15} 
\multicolumn{1}{|c|}{Deployed} & 22 464                                                & 42 519                                                & 3 078                                                 & 5 282                                                 & \multicolumn{1}{c|}{} & Grouped           & 40 865                                                & 22 478                                                & 6 879                                                 & 1 481                                                 & 148                                                & 63 343                                                  & 0                                                  & 8 360                                                   \\ \cline{1-5} \cline{7-15} 
\multicolumn{1}{|c|}{All dep.} & 42 590                                                & 65 753                                                & 3 868                                                 & 6 788                                                 & \multicolumn{1}{c|}{} & Not grouped       & 22 464                                                & 42 519                                                & 3 078                                                 & 5 282                                                 & \multicolumn{2}{c|}{NA}                                                                                      & \multicolumn{2}{c|}{NA}                                                                                      \\ \cline{1-5} \cline{7-15} 
\end{tabular}
\vspace{-10pt}
\end{table*}

%---------------------------------------------------------------------------------------
\begin{figure}[t]
% \vspace{-10pt}
\includegraphics[width=0.9\columnwidth, keepaspectratio]{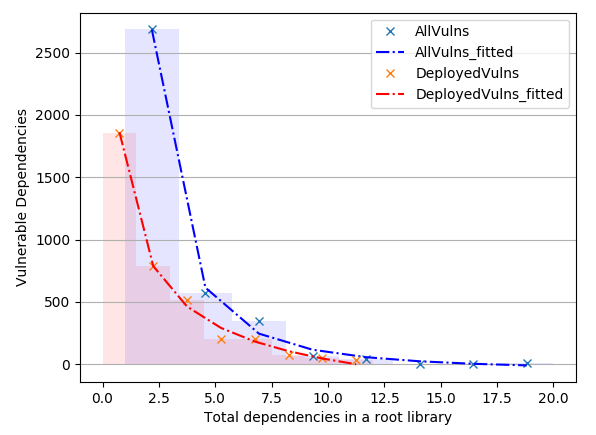}
\longcaption{\columnwidth}{Significant amount of vulnerabilities are coming from non-deployed dependencies. However, these vulnerabilities are not exploitable, and therefore, do not introduce any danger to the analyzed library instances.}
\caption{All vs Non-deployed vulnerable dependencies}
\vspace*{-\baselineskip}
\label{fig:all:nondeployed}
\end{figure}
%---------------------------------------------------------------------------------------

To answer RQ1 we collected both direct and transitive dependencies without applying other simplification steps. The left part of Table~\ref{table:cure:effect} shows the total amount of both vulnerable and safe dependencies in our sample. We found that non-deployed dependencies represent 45\% of direct and 34\% of transitive dependencies, wherein 22\% of transitive and 20\% of direct vulnerable dependencies are non-deployed.

Figure~\ref{fig:all:nondeployed} shows the distributions of the total number of vulnerable dependencies and the number of vulnerable dependencies of an analyzed library that are actually deployed. We observe, that those distributions are different: the number of actually deployed dependencies is significantly smaller than the total number of dependencies in a library (p-value=$1.467*10^{-190}$, Wilcoxon test). In some cases only 22\% of vulnerable dependencies are released with the library, while the majority of the analyzed libraries in our sample have up to 12 vulnerable dependencies, three of which are non-deployed (~25\%).

We observe that non-deployed dependencies are a significant share of vulnerable libraries: \textbf{every fifth dependency affected by a known vulnerability is non-deployed, and does not bring any danger to the analyzed library}. Hence, they should be excluded (or separately marked) from the results of a dependency study.

%---------------------------------------------------------------------------------------
\paragraph{RQ2: Who is responsible for vulnerable dependencies?}

%---------------------------------------------------------------------------------------
\begin{figure}[t]
% \vspace{-10pt}
\includegraphics[width=0.9\columnwidth, keepaspectratio]{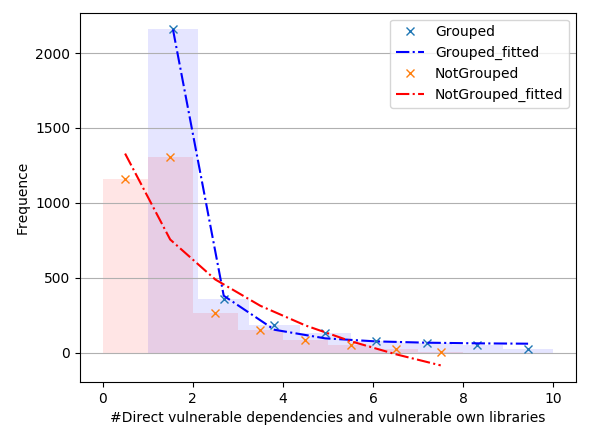}
\longcaption{\columnwidth}{The grouping step allows us to report many vulnerable cases that software developers of the analyzed libraries are actually in direct control of, since those vulnerabilities are present in either direct dependencies or own libraries of the analyzed projects.}
\caption{The cases when developers of analyzed libraries are actually responsible for fixing vulnerable dependencies
}
\label{fig:hist:grouped:notgrouped}
\vspace{-20pt}
\end{figure}
%---------------------------------------------------------------------------------------

% As it was discussed in Section~\ref{sec:sec:own:third}, developers are responsible for updating own libraries and direct dependencies of their software projects.
To make an application safe, its developers need to be sure that they address all the 
vulnerable dependencies. Direct 
dependencies and own libraries of a software project are within the full control of 
its developers but fixing vulnerabilities in transitive dependencies may 
require opening the ``black-box'' approach for dependency management, and may be significantly more expensive. 
%\todo[inline]{.}
%\todo[inline]{IP: What do you think about such a fix? BTW, maybe we should use the
%term ``control zone'', instead of ``responsibility''?}
To answer the RQ2 we need to determine the following:
\begin{compactitem}
    \item the difference between the ``true'' number of own libraries and direct dependencies in the dependency trees of analyzed libraries;
    \item the actual number of vulnerable dependencies that developers of analyzed libraries are responsible for fixing.
\end{compactitem}
To do this, firstly, we filtered out non-deployed dependencies. Then we identified the own dependencies and compared the number of direct dependencies before and after the grouping procedure.

The effect of the grouping procedure on the aggregated numbers of dependencies for the studied library sample is shown in the right part of Table~\ref{table:cure:effect}. We found that paths of 148 (out of 10905) analyzed library instances include own dependencies of the analyzed projects. However, those own dependencies are not affected by known vulnerabilities. This most probably reflects the quality of the selected OSS library sample: the OSS libraries used by SAP belong to well-organized software projects. However, these own dependencies may still introduce some noise while creating a bill of materials to plan allocation of development resources.

Besides own dependencies, software developers are also responsible for fixing their direct dependencies (See section~\ref{sec:sec:own:third} for detailed discussion). After the grouping step we observe the surprising increase of the number of direct dependencies by as much as 87\%. This most likely happens because the dependency grouping procedure shortens the dependency paths (by grouping dependencies belonging to same projects), and therefore, it reduces the appalling feeling of an unmanageable `dependency hell'.

Figure~\ref{fig:hist:grouped:notgrouped} shows the difference between the distributions of the number of vulnerable dependencies that developers of analyzed libraries are in direct control of (own libraries and direct dependencies) before and after application of the grouping step (the difference is statistically significant, p-value=$1.648*10^{-285}$). The dependency grouping allows us to reveal up to eight additional vulnerable dependencies under the direct control of the developers of an analyzed library instance.

We observe that developers of the analyzed libraries \emph{could} fix (either by directly correcting a bug in the library instances belonging to their project or by updating direct dependencies to newer versions) the major part of vulnerable dependencies, because these are under their responsibility.
\textbf{Without the dependency grouping, it may seem that developers of the analyzed libraries have direct control of only 37\% of the vulnerable dependencies, while in reality they are responsible for fixing 82\% of the deployed vulnerable dependencies}.

%---------------------------------------------------------------------------------------
\paragraph{RQ3: How many direct dependencies can be actually fixed?}

\begin{table*}[t]
\caption{The effect of halted dependencies (RQ3)}
\label{table:halted}
\vspace{-10pt}
\longcaption{\textwidth}{14 \% of dependencies of the analyzed library instances are halted, while 1 \% of them are affected by known vulnerabilities. Moreover, the right part of the table shows, that direct halted dependencies introduced 7 vulnerable dependencies transitively.}
\begin{tabular}{c|c|c|c|c|cc|c|c|c|c|c|c|}
\cline{2-5} \cline{8-13}
 & \multicolumn{2}{c|}{Not vuln} & \multicolumn{2}{c|}{Vuln} &  &  & \multicolumn{3}{c|}{Not vuln} & \multicolumn{3}{c|}{Vuln} \\ \cline{2-5} \cline{8-13} 
 & Direct & Transitive & Direct & Transitive &  &  & Halted & Outdated & Up-to-date & Halted & Outdated & Up-to-date \\ \cline{1-5} \cline{7-13} 
\multicolumn{1}{|c|}{Halted} & 5 369 & 3 678 & 69 & 5 & \multicolumn{1}{c|}{} & \begin{tabular}[c]{@{}c@{}}Transitive\\ via halted\end{tabular} & 187 & 378 & 0 & 0 & 7 & 0 \\ \cline{1-5} \cline{7-13} 
\multicolumn{1}{|c|}{All dep.} & 40 865 & 22 478 & 6 879 & 1 481 & \multicolumn{1}{c|}{} & All dep. & 9 047 & 54 836 & 1 731 & 74 & 8 286 & 0 \\ \cline{1-5} \cline{7-13} 
\end{tabular}
\vspace{-20pt}
\end{table*}

To answer RQ3 we considered only deployed dependencies, grouped according to the software projects
they belong to. We found that 13\% of the overall number of direct dependencies and 16\% of transitive dependencies are halted. Some of them 
% \as{When reporting numbers (non-percentages) always indicate ``out of X'' to make it clearer}
(69 direct and 5 transitive out of 9047 halted dependencies) are affected by known security vulnerabilities. Although this number is not big, each case of a halted dependency is very important. Such dependencies do not have a fixed version, and therefore, a costly mitigation is needed to fix such vulnerabilities.

Additionally, 
within the sample of 10905 analyzed libraries, we found five library instances that have transitive vulnerable dependencies via a halted direct dependency. All these dependencies are outdated and there exist safe versions of them. However, these safe versions would not be adopted by halted libraries, and therefore, developers of analyzed libraries have to apply a non-trivial mitigation strategy: to artificially convert those dependencies into direct dependencies of their libraries. 

The proposed methodology allowed us to identify that \textbf{14\% of the dependencies in our sample are halted, while 1\% of them are affected by known security vulnerabilities}. Moreover, \textbf{direct halted dependencies also transitively introduced 565 dependencies, 7 of which are vulnerable}. All these vulnerabilities require specific costly mitigation strategies.
% \todo{OPTIONAL: Do we have the information about the total? E.g. can we say something like seven library instances  over XXX that have transitive vulnerable..... }

\noindent\paragraph{\textbf{Effect of the proposed methodology on an individual software library}}

% To underline the importance of the proposed methodology, we show its impact being applied to a typical industrial software library.
To identify a typical industrial software library, we have extracted the number of direct dependencies for each SAP software library in the proprietary Nexus repository. We assume that the number of direct OSS dependencies in a typical industrial library is equal to the mean number of direct dependencies that SAP projects have, which we found to be equal 12.

Then we have artificially constructed dependency trees for 100 software projects:
\begin{compactenum}
    \item From the overall sample of analyzed libraries we randomly select 12 libraries
    \item For each selected library in the step 1, we randomly pick its version
    \item We calculate the difference between the results received according to the ``standard'' dependency study methodology (used, for example, in~\cite{kula2017ese}) and the proposed methodology
    \item We repeat steps 1--3 100 times to receive the data for the specified number of simulated projects
\end{compactenum}

\begin{figure*}[!t]
\subfloat[RQ1 for a typical industrial library]{
\includegraphics[width=0.85\columnwidth, keepaspectratio]{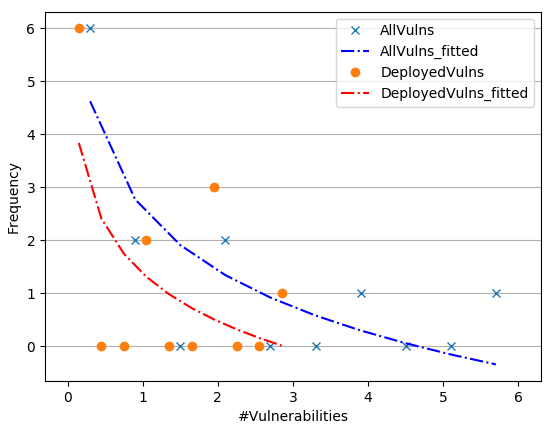}
\label{fig:proj:rq1}
}
%-------
\subfloat[RQ2 for a typical industrial library]{
\includegraphics[width=0.85\columnwidth, keepaspectratio]{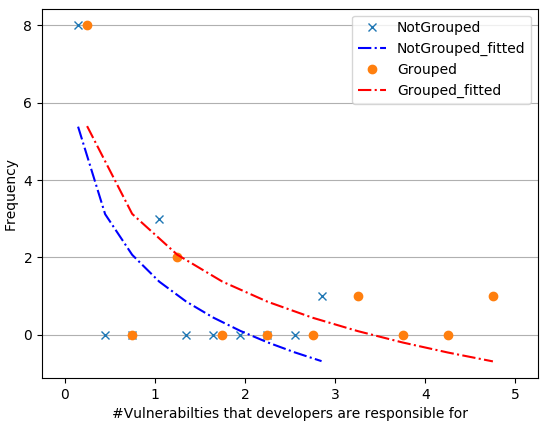}
\label{fig:proj:rq2}
}

\longcaption{\textwidth}{The proposed methodology has a positive impact on the dependency analysis results of a typical industrial library, since it allows us to distinguish deployed vulnerable dependencies from all vulnerable dependencies for a typical industrial library (Figure~\ref{fig:proj:rq1}), as well as, to report the actual number of dependencies in direct control of developers of industrial libraries (Figure~\ref{fig:proj:rq2}).}
\vspace{-10pt}
\caption{Effect of the proposed methodology for a typical industrial library}
\label{fig:proj:effect}
\vspace{-10pt}
\end{figure*}

Figure~\ref{fig:proj:effect} shows the effect of the proposed methodology for a typical industrial library. We observe that the number of deployed vulnerable dependencies is always lower, than the total number of vulnerable dependencies (Figure~\ref{fig:proj:rq1}). At the same time we see that the proposed methodology allows us to distinguish additional dependencies that developers of the simulated industrial libraries are responsible for (own and direct dependencies). Additionally, we found that an average library in our simulation have a 9,5 \% chance to have a vulnerable halted dependency ($\sigma=0.252$) with a maximum number of 2 vulnerable halted dependencies.

Hence, we can conclude that \textbf{the proposed methodology has a positive impact on the correct resolution of dependency analysis results of a single industrial library}.

%\todo[inline]{IP: Think how to include a comparison of our results with results received according to OWASP Dependency check}

%---------------------------------------------------------------------------------------

\section{Implications on Industrial Practice}
\label{sec:ind:implication}
In an industrial setting, the practical negative impact of using an \emph{inadequate} measurement method can be substantial. Ensuring a healthy supply chain of third-party dependencies (of which the large majority is OSS)
is a continuing effort that spans the development and the operational phases of a product lifetime.

As part of SAP's secure development life-cycle, all development projects go through several validation steps and
each single finding has to be audited, assessed, and mitigated.
After the product is released to customers, and for its entire
operational lifetime, its own security and the security of its third-party dependencies are continuously monitored. When a vulnerability is detected in one of the dependencies, timely mitigations need to be developed and deployed to all affected systems. In the case of OSS dependencies,
these mitigations may consist of dependency updates, or in ad-hoc fixes in the product that relies on the affected
library or in the dependency itself (through a company-internal fork that can be temporary or persistent). When the product portfolio of a company includes thousands of products, whose support period can extend to decades,
wrong assessments lead to inadequate risk management and inefficient allocation of resources, which ultimately translate to increased chances of security incidents and financial loss.

The distinction between deployed and non-deployed components allows quick and reliable pre-filtering of not exploitable vulnerable dependencies, since they are not part of the deployed product.
From our analysis of a sample of over 550 OSS libraries used by SAP projects, as many as 20\% of all the dependencies are \emph{non-deployed}. Any metrics reporting the ``danger'' of using OSS libraries that do not discriminate between those two classes would lead to a wrong allocation of costly development and audit resources.

The granularity at which dependencies are analyzed and the reliability with which vulnerabilities affecting them are detected are essential to obtaining a meaningful
view of the (security) health of the dependencies of a project. 
Approaches that use imprecise vulnerability detection methods and that ignore the interdependencies among the individual nodes of the dependency tree yield a distorted view, which requires
tedious, manual reviews to be correctly interpreted and that cause precious resources to be wasted. Failing to group dependency nodes that belong to the same group (e.g., to the same OSS project), and that are updated together, makes the update of certain libraries appear more problematic than it is. The vulnerability may affect a node that is deep in the dependency tree, while the node that the application developer would need to update might be much shallower (e.g., it could even be a direct dependency). More in general, imprecise approaches to vulnerability management undermine the trust of developers on automated  analysis because the dependencies identified as problematic do not correspond to those that must be actually acted upon to address the reported issues. As a consequence, despite the promises of automation, considerable additional human effort and expert judgment is required to determine the appropriate mitigation strategy.

Finally, determining precisely whether a dependency could be upgraded to a non-vulnerable version or not (because such a version does not exist, and perhaps will never exist, if the dependency is no longer maintained) is the key to choosing the correct mitigation strategy.
Addressing vulnerabilities in OSS components that are alive, but for which a fixed release does not exist \emph{yet}, requires to act fast, so that an emergency solution
can be rolled-out as fast as possible to all customers. Being temporary and urgent, such mitigation might not be optimal. An upgrade to a non-vulnerable version of the dependency will eventually be done.
Conversely, if a vulnerability affects a dependency that is no longer maintained, fixing the code of the dependency would effectively mean creating a company-internal fork, whose long-term support could require substantial additional investments and maintenance effort.

\section{Related Work}
\label{sec:relwork}
% \todo[inline]{What about \citep{kula2017ese,bavota2015ese,cox2015icse}?}
% \todo[inline]{IP: \cite{bavota2015ese} does not study dependencies with known vulnerabilities. Therefore, I do not include the reference to this paper in this section. However, I think, the paper can be used in introduction section to motivate our study.}

\subsection{Dependency Studies}

Williams and Dabirsiaghi~\citep{williams2012unfortunate} report that 26\% of open-source libraries downloaded by organizations from Maven Central to have known vulnerabilities and average software projects to contain at least one vulnerable dependency. The authors refer to a lack of meaningful controls of the components used in the proprietary software projects as a possible reason for such a high number of usage of vulnerable dependencies.

Hejderup \cite{hejderup2015dependencies} studied the \textsf{npm} registry of JavaScript modules and found that one-third of all modules use vulnerable dependencies. Besides the lack of awareness of developers, the study suggests context usage of a module and breaking changes to be the possible reasons for not fixing vulnerable dependencies. However, the authors did not distinguish deployed and non-deployed dependencies, hence the results may be reported for low-priority libraries.

The first large scale study of JavaScript open source projects was done by Lauinger et al.~\cite{lauinger2017thou}. The authors underline the finding that transitive dependencies of a project are more likely to be vulnerable, since developers (i) may not be aware about their existence and (ii) they have less control on them. However, this relation between direct and transitive dependencies seems to be specific for the JavaScript environment, since it allows different versions of the same dependency to be included several times. Moreover, the authors say that main sources of transitive dependencies in the web sites are advertisement, tracking or social widget code, security side of which is not very well maintained. The other dependency management system may not have such problems by design. For example, Maven allows a project to use only one version of a dependency, while open-source Java projects typically do not include advertisement or social widget contents.

The authors investigate the relation between outdated dependencies and dependencies with known vulnerabilities. However, Cox et al.~\cite{cox2015icse} extract dependencies from project \textit{pom.xml} files, which means that the study report results only for direct dependencies and do not apply Maven version resolution procedure. Although the latter does not have a high impact while working with direct dependencies, it may introduce errors when transitive dependencies are involved.
Moreover, the study might include low-priority findings, since it does not explicitly mention that non-deployed dependencies were filtered out.
We propose to use Maven resolution procedure and consider results for both direct and transitive dependencies. Cox et al.~\cite{cox2015icse} rely on name-based matching of CVEs onto library dependencies, which may have a high number of false positives~\cite{cadariu2015tracking}. Instead, we propose a precise matching approach, which relies on code-level matching.
%1.	They say that bad-ranked systems (which use many outdated dependencies) are 4 times more likely to have dependencies with known vulnerabilities, but according to their table, the 5-star system (the most ‘fresh’ system) has the same 33\% chance of having outdated dependency as 1-star system -> we should propose a more meaningful way to correlate outdated dependencies with vulnerable dependencies

Kula et al.~\cite{kula2017ese}
% studied if developers update dependencies of their projects. They
report 81,5\% of the studied projects to have outdated dependencies, and 69\% of the project owners to be unaware of vulnerable dependencies in their projects. Although the authors provide a thorough insight into developers' motivation of keeping dependencies outdated, the study uses manual analysis to map security advisories onto affected project versions, and therefore, cannot be easily applied to a large number of software projects (also, the study provides insights for only nine versions  of three libraries). For the study of dependencies with known vulnerabilities Kula et al.~\cite{kula2017ese} used security advisories for just five CVEs of two types - Denial of Service and ``man in the middle''. Hence, the results of the study might not cover 
important aspects of the problem of outdated dependencies.
% Moreover, the authors did not use latest versions of software libraries, because they did not allow the study of library migration trends.
Also, as the reported developer comments reveal, the study did not filter out non-deployed dependencies and did not consider grouping dependencies by projects.

% The threats underlined in the related works inspired us to come up with the methodological approach for the study of library dependencies. There are several works that propose various improvements to the dependency management process, however, each of dependency studies starts the investigation from its own side, hence, revealing a part of the general problem. Instead, the dependency management is a complicated, yet holistic, process, and therefore, should be addressed thoroughly.

\begin{table}[t]
\caption{Aspects considered in the related works}
\label{table:relwork:rqs}
\vspace{-10pt}
\footnotesize
\begin{tabular}{|c|c|c|c|c|}
\hline
Related work & \begin{tabular}[c]{@{}c@{}}RQ1: only\\ deployed?\end{tabular} & \begin{tabular}[c]{@{}c@{}}RQ2:\\ grouped?\end{tabular} & \begin{tabular}[c]{@{}c@{}}RQ3:\\ halted deps\end{tabular} & \begin{tabular}[c]{@{}c@{}}Vuln.\\ matching\end{tabular} \\ \hline
\begin{tabular}[c]{@{}c@{}}Williams and\\ Dabirsiaghi~\cite{williams2012unfortunate}\end{tabular} & No & No & No & Name-based \\ \hline
Hejderup~\cite{hejderup2015dependencies} & No & NA & No & Name-based \\ \hline
Lauinger et al.~\cite{lauinger2017thou} & Yes & NA & No & Manual \\ \hline
Cox et al.~\cite{cox2015icse} & No & No & No & \begin{tabular}[c]{@{}c@{}}Name-based+\\ manual\end{tabular} \\ \hline
Kula et al.~\cite{kula2017ese} & No & No & No & Manual \\ \hline
\begin{tabular}[c]{@{}c@{}}OWASP Dep Check\end{tabular} & NA & NA & NA & Name-based \\ \hline
Cadariu et al.~\cite{cadariu2015tracking} & No & No & No & Name-based \\ \hline
Alqahtani et al.~\cite{alqahtani2016tracing} & No & No & No & Semantic-web \\ \hline
Ponta et al.~\cite{ponta2018beyond} & NA & NA & NA & Code-based \\ \hline
\end{tabular}
\vspace{-20pt}
\end{table}

%==============================================================================
\subsection{Identification of Vulnerable Dependencies}
\label{sec:background:approaches}

OWASP Dependency Check\footnote{\url{https://www.owasp.org/index.php/OWASP_Dependency_Check}} is a tool, which provides the functionality to automatically extract a list of project dependencies and check if this list contains any libraries with known security vulnerabilities. The tool allows automatic matching of a library with an associated CVE by comparing the name of a library with a CPE version indicated in the description of a vulnerability (CVE) in NVD. Although such approach has high performance, it fully relies on the information present in the NVD.

Cadariu et al.~\cite{cadariu2015tracking} enhanced the OWASP Dependency Check tool to create a Vulnerability Alert Service (VAS) to provide the information about vulnerable dependencies used by clients of the Software Improvement Group (SIG). However, the authors discovered that the matching mechanism based on comparing library names with CPEs yields many false positives.
Moreover, at the time of publication of~\cite{cadariu2015tracking} VAS was capable only to provide information regarding direct dependencies, while vulnerabilities may be also introduced via transitive dependencies~\cite{hejderup2015dependencies}.

Alqahtani et al.~\cite{alqahtani2016tracing} used a semantic-web approach for mapping CVE descriptions from NVD database to the corresponding Maven library identifiers. However, the precision of the approach is 5\% lower when compared to OWASP Dependency Check (and consequently to VAS). Hence, the results reported in~\cite{alqahtani2016tracing} may provide inaccurate estimation of the number of vulnerable dependencies in the open-source projects being affected by both FP and FN.
% Moreover, Alqahtani et al. were checking every artifact in Maven Central, hence, they may potentially report findings for unused libraries. Instead, for our study we used the latest released versions of popular OSS projects, which eliminates the threat of reporting results for unused libraries.

We rely on the works from Plate et al.~\cite{plate2015impact} and Ponta et al.~\cite{ponta2018beyond}, who propose a precise approach to use the patch-based mapping of vulnerabilities onto the affected components (see Section~\ref{sec:vuln:identification}).

\section{Threats to Validity}
\label{sec:threats}
Threats to \textit{internal validity} concern the external factors not considered in our study:

% \as{This is outdated right? it does not cover the fact that the selection of subjects does not come from Github but from SAP, and that we do not only cover Maven projects, but any Java project (as long as it is in Maven Central, that is all of them)}
% \todo[inline]{IP: I have updated the Threats to Validity section}

\textit{The selection of OSS libraries is based on the number of usages from within \SAP.} Such selection criterion may yield a sample not representative of what libraries are most relevant for other industrial companies or OSS developers. To check the popularity of the studied libraries within the OSS community, we obtained the information about library usages from MVNRepository and the number of OSS contributors that claimed to use the selected libraries from BlackDuck Openhub\footnote{https://www.openhub.net/}. The results obtained from both sources suggested us that selected libraries are popular within the OSS developers. Since \SAP is a large multinational software development company with a significant number of Java projects, we believe that the threat of industrial non-representativeness is minimal.

% Other possible criteria may be the number of project contributors, the number of modules in a project, or the number of downloads from the Maven Central Repository. However, in this study we want to raise the awareness about the false sense of security that goes with the latest released versions of OSS projects, rather than make any claims regarding security of OSS projects in general.

\textit{The vulnerability database used for our study may not cover all known vulnerabilities.} To minimize this threat \SAP conducted an internal study of the vulnerability dataset, which concluded that it covers 90\% of all NVD vulnerabilities reported for OSS projects developed in Java. The coverage is closer to 100\% when considering the OSS projects most relevant for \SAP.
% Moreover, while analyzing software dependencies during this study, we kept track of unknown GAVs and did not find any not being registered in the database.
Hence, we believe that this threat has minimal influence on the results of our analysis.

% \todo[inline]{HP: The before-last sentence is not clear to me: I understand that we did not come across GAVs unknown to Vulas, but what does that mean for the coverage of the Vulas vulnerability DB? This are different things...}
% \todo[inline]{IP: I removed that sentence}

% \textit{We do not compare project release and CVE publishing dates.} It may happen, that for some projects the information of a vulnerability in a dependency had not been published by the release date. However, we believe that this threat is minimal, since we studied latest available released versions of software projects. Moreover, most of found CVEs were published before 2017.

Threats to \textit{external validity} concern the generalization of results:

%\textit{We analyze a subset of most popular Maven based Java projects.} In this study we present the results for only a subset of all Maven based Java projects, because we want to raise the awareness of both OSS and industry communities on the problem of having vulnerable dependencies. However, the methodology presented in this paper can be easily adapted for a large-scale study.

\textit{Currently we support only Maven based projects.} We used Maven, because it provides very comfortable way to handle dependency management and is wildly used within both OSS and commercial projects. Clearly, dependency analysis can be enlarged to other build automation systems, like Ant or Gradle. Although our tool depends significantly on Maven, the methodology itself may be applied to study projects using other build automation systems.

\textit{We use Maven groupIds as an approximation for a project.} This may potentially lead to an incorrect grouping of libraries because some projects may use the same cross-project groupIds, or conversely, different groupIds to identify their components. The former threat has a minimal impact, since the Maven naming convention of assigning different group identifiers to distinct projects is quite well established. We observed the latter case for test or example libraries, e.g., \code{org.apache.activemq} has a subgroup \code{org.apache.activemq.tooling}. We considered two groupIds as equal if one of the two includes the other groupId (as in the \code{activemq} example).
The projects that cannot be distinguished only by groupId could be distinguished using additional atributes, such as \textit{Repository}, \textit{ProjectID}, and others (which might be specific to certain language ecosystems).

% For our study, the distinguishing power of Maven groupId is enough, although for a large-scale study the more accurate identification of a project is required.

% Threats to reproducibility concern the possibility of replication of results:

% \textit{We do not share the vulnerability database used in this study, since it is a proprietary database of SAP.} However, an interested reader may refer to the NVD, or any other source of information that allows identification of patches corresponding to vulnerability fixes. Then the patch matching approach to reconstruct the vulnerability database may be applied (see Section~\ref{sec:vuln:identification}). To ease the replication of our study we plan to share the tool we developed to collect and resolve software dependencies under an open-source license.

% \todo[inline]{IP: Do you think, if we need to mention here that we do not report whether there exists a safe version of a dependency?}

% \textit{We do not report whether there always exist a safe version of a dependency.} In some cases a vulnerable dependency may be outdated, but there is no vulnerability fix provided yet. Although we do not report results for this scenario, the treatment should be similar to the case of halted dependencies. So, our methodology may be easily adopted to also report the cases, where a certain dependency does not have a fixed version.

%\section{Replicability and Reliability}
%\label{sec:replicability}
%\input{replicability}

\section{Conclusions and Future Work}
\label{sec:conclusions}
In this paper we have proposed a methodology for reliable measurement of vulnerable dependencies in OSS libraries. To demonstrate our methodology, we selected 200 most used OSS Maven based libraries from within \SAP. However, the concepts underlined in our methodology apply to any dependency management system. 

To perform the analysis we have built a tool that leverages the functionality of Apache Maven to extract the library dependencies and applies code-level matching approach to identify the known vulnerabilities affecting them. We have also performed several post-processing steps, such as (i) filtering non-deployed dependencies, (ii) grouping dependencies on their belonging to software projects, and (iii) determining whether a certain dependency is halted.

The results of our study demonstrate that all the suggested post-processing steps have a positive impact:
\begin{compactitem}
\item every fifth dependency affected by a known vulnerability is non-deployed, hence our methodology allows reported results of a dependency study to be free from a significant number of vulnerable dependencies that do not introduce any harm to the analyzed libraries;
\item the grouping step of the proposed methodology allows us to reveal 82\% (45\% more comparing to a regular approach) of vulnerable dependencies the developers of the analyzed libraries are responsible for fixing;
\item the results of the dependency study suggest that 14\% of the total number of dependencies are halted, and therefore, do not receive updates (including security fixes). Such dependencies should be used with caution, since mitigations of their bugs and bugs of their dependencies are costly;
\item the library simulation shows that the proposed approach has a positive impact on the correct resolution of dependency analysis results of a single industrial library.
\end{compactitem}

As future directions of our research we take the following steps:
\begin{compactitem}
\item to investigate the situation outside of the Maven ecosystem, for example
targeting \textit{npm} or \textit{pip}.
\item to extend this study to analyze all the existing libraries in Maven Central;
\item to identify a precise model for automatic identification of whether a certain library is halted;
\item to complement the existing studies on the reasons why developers do not update dependencies with an investigation of developers' behavior with regard to security-related updates.
% \item to study 
% the reaction of developers when they are first made aware
% of the existence of vulnerable dependencies
% the reactions of library developers, to
% whom we provide the detailed reports of dependency studies.
\end{compactitem}

\bibliography{biblio}
\bibliographystyle{abbrv}

\end{document}